\documentclass[journal]{rmaa}
\usepackage{caption}

\def\fov{FoV }
\def\RRab{RRab }
\def\RRc{RRc }

\usepackage{xcolor}

\begin{document}

\title{Variable stars in NGC 4147 revisited: RR Lyrae stars new CCD \emph{VI}
photometry and physical parameters}

% Note the difference betwen `and' and `\and'. The latter appears on a 
% line of its own. 
  %% note that the `\and' command is necessary to put the `and' on its
  %% own line
\author{
  A. Arellano Ferro,\altaffilmark{1}
  F.C. Rojas Galindo,\altaffilmark{2}
  S. Muneer,\altaffilmark{3}
  Sunetra Giridhar\altaffilmark{3}
}

\altaffiltext{1}{Instituto de Astronom\1a, Universidad Nacional Aut\'onoma de
M\'exico, M\'exico.}
\altaffiltext{2}{Instituto de F\'isica, Universidad de Antioquia, Colombia}
\altaffiltext{3}{Indian Institute od Astrophysics, Bangalore, India}

\fulladdresses{
\item A. Arellano Ferro: Instituto de Astronom\1a, Universidad Nacional
Aut\'onoma de M\'exico, Apdo. Postal 70-264, Ciudad de M\'exico, CP 04510, 
M\'exico. (armando@astro.unam.mx)

}

\shortauthor{Arellano Ferro et al.}
\shorttitle{NGC 4147 revisited}

\SetYear{2018}
\ReceivedDate{January 2018}
\AcceptedDate{------} 

\resumen{Los par\'ametros f\'isicos en estrellas RR Lyrae en NGC 4147 se calcularon
por medio de la descomposici\'on de Fourier de sus curvas de luz, empleando
calibraciones y puntos cero bien establecidos. Las estrellas RRc indican los valores
medios de metalicidad y distancia al c\'umulo de [Fe/H]$_{ZW}=-1.72\pm0.15$ and
$19.05\pm0.46$ kpc respectivamente. La estrella V18, cuya variabilidad hab\'ia sido
puesta en duda, es confirmada ahora como una variable del tipo SR con un periodo
de 24.8d y pudo ser empleada para estimar la distancia al c\'umulo de manera
independiente. Se observa que las estrellas RRab y RRc comparten la zona inter-modo en
la zona de inestabilidad.
El c\'umulo puede clasificarse como del tipo intermedio de Oosterhoff. La
estructura de su rama horizontal y su metalicidad favorecen la interpretaci\'on de su
origen extragal\'actico. En el plano $M_V-$[Fe/H] el c\'umulo sigue la tendencia
de los c\'umulos tipo Oo~I, definida por sus estrellas tipo RRc. }

\abstract{We have calculated the physical parameters of the RR Lyrae stars in
the globular cluster NGC~4147 via the Fourier decomposition of their light curves,
using new data and well-established semi-empirical calibrations and zero points.
The mean metallicity and distance estimated using the RRc stars are
[Fe/H]$_{ZW}=-1.72\pm0.15$ and $19.05\pm0.46$ kpc respectively. The star V18, whose
variability has been previously in
dispute, is confirmed to be a variable of the SR type with a period of about
24.8d,
and it has been used to get and independent distance estimate of the cluster. It is
observed that the RRab and RRc stars do not share
the inter-mode region in the horizontal branch. The cluster can be classified as of
intermediate Oosterhoff type. Its horizontal branch structure and metallicity make
a good case for extragalactic origin.  It follows the distribution of Oo~I type
globular clusters in the $M_V-$[Fe/H] plane, as depicted from the RRc stars.} 


\maketitle

\section{Introduction}
\label{sec:intro}

The globular cluster NGC 4147 is a faint and distant cluster some 19 kpc from
the Sun and 21 kpc from the Galactic Center ($\alpha = 12^{\mbox{\scriptsize h}}
10^{\mbox{\scriptsize m}} 06.^{\mbox{\scriptsize s}}$, $\delta =
+18^{\mbox{\scriptsize o}} 32{\mbox{\scriptsize '}} 31.8{\mbox{\scriptsize "}}$,
J2000; $l = 252.84^{\mbox{\scriptsize o}}$, $b = +77.18^{\mbox{\scriptsize o}}$).
It is one of the globular clusters immersed in the Saggitarius Stream 
and it has been suggested that originally it may have been a cluster member of the 
Sagittarius dSph galaxy (Bellazzini et al. 2003a, 2003b).
Its classification as an Oosterhoff type I (Oo~I) (Castellani \& Quarta 1987) was
controversial given its low metal abundance, blue horizontal branch (HB) and the
large fraction of RRc type stars among the RR Lyrae population. This
classification was
confirmed by Arellano Ferro et al. 2004 (hereinafter AF04) from the revised
periods of the RR Lyrae stars and by Stetson, Catelan \& Smith (2005) (hereinafter
SCS05) from the RR Lyrae distribution in the amplitude-period plane.
It is probably the most metal poor globular cluster among the Oosterhoff type I
clusters. However, considerations on the HB structure and amplitude-period
distribution of the RRc stars, trigger the possibility that the clusters is of an
Oosterhoff intermediate type (Oo-Int) and of extragalactic origin, as shall be
discussed later in this paper.

The cluster variable stars were studied by AF04 
in the $V$ and $R$ band passes and the light curves of the
RR Lyrae were Fourier decomposed to estimate the mean metallicity and distance to the
cluster. However, since then, we have improved upon several procedures related with
the data reductions (Bramich et al.\ 2013) and transformations to the standard system,
on the
usage of the semi-empirical calibrations and their zero-points to estimate physical
properties of RRab and RRc stars in a very homogeneous way (Arellano Ferro et al.
2013; 2017), and have expanded the critical analysis of the RR Lyrae
distribution on the colour magnitude diagram (CMD) to a family of
clusters of both Oosterhoff types I and II (Oo~I and Oo~II) (e.g. Yepez et al. 2018). 
A Fourier decomposition of both RRab and RRc stars was also performed by SCS05. In
both the above studies, it became clear that the light curves morphology in all the
five known RRab is peculiar, due to the observed amplitude modulations of their
light curves (AF04) and consequently displaying a large parameter morphology $D_m$
(SCS05), which reveals the light curve as unsuitable for metallicity calibration.
This is probably the reason why, both AF04 and SCS05, concluded on a much
higher metallicity for the cluster. Among the RRc
stars, numerous anomalies in the rather long period variables (V11, V13, V16, V17)
were encountered in the above papers, such as erratic phasing and amplitude
modulations
that may be connected with the presence of secondary frequencies and/or the Blazhko
effect. Thus, a reconsideration of the stellar physical quantities as obtained from
the Fourier light curve decomposition is in order.
For all the above reasons, the cluster was observed in 2012  with the aim of enriching
the data collection and time-base for a proper comparison with previous discussions
and to set the metallicity and distance to the cluster in the homogeneous scales
of a family of some 24 clusters presented by Arellano Ferro et al. (2017) from the
Fourier decomposition approach and their inference in the $M_V -$[Fe/H] relation.

The paper is organised as follows; we describe our observations and data reductions as
well as the
transformation to the Johnson-Kron-Cousins photometric system in $\S$ 2, we perform
the
identification of the known variables, estimate the cluster reddening in $\S$ 3, we
calculate the physical parameters via the Fourier decomposition for RR Lyrae stars in
$\S$ 4, discuss the peculiarities of some individual stars in $\S$ 5, comment on the
distance to the cluster via several methods in $\S$ 6,
discuss the CMD and the isochrone and zero age horizontal branch (ZAHB) fitting
in $\S$ 7, the Oosterhoff type of NGC~4147, the cluster position in the Oosterhoff
gap and hence its possible extragalactic origin are the topics treated in $\S$ 8,
we present NGC~4147 in the perspective of the $M_V$-[Fe/H] relation
determined from RRc stars in a family of globular clusters in $\S$ 9, and finally we
summarise our results in $\S$ 10.

\section{Observations and reductions}
\label{sec:ObserRed}

\begin{table}[t]
\footnotesize
\caption{The distribution of observations of NGC~4147.
Columns $N_{V}$ and $N_{I}$ give the number of images taken with the $V$ and $I$
filters respectively. Columns $\MakeLowercase{t}_{V}$ and $\MakeLowercase{t}_{I}$
provide the range of exposure times. In the last column the average seeing is listed.}
\centering
\begin{tabular}{lccccc}
\hline
Date  &  $N_{V}$ & $t_{V}$ (s) & $N_{I}$ &$t_{I}$ (s)&Avg seeing (") \\
\hline
 2012-02-05 & 35 & 140-350 & 18 & 30-150   & 2.5\\
 2012-02-28 & 20 & 170 250 & 19 & 50-120   & 1.7\\
 2012-02-29 & 12 & 300-450 & 11 & 90-250  & 3.0\\
 2012-03-01 & 35 & 240-350 & 35 &  85-120 & 2.4\\
 2012-03-02 & 18 & 150-300 & 19 & 60-150 & 2.2\\
 2012-04-11 & 7 & 250-300 & 7 & 110-150 & 2.5\\
 2012-04-28 & 11 & 300-350 &  2 & 100-150 & 2.7\\
 2012-04-29 & 15 & 220-300 & 10 & 100-200 & 2.2\\
\hline
Total:   & 153&    &  121  &    &\\
\hline
\end{tabular}
\label{tab:observations}
\end{table}

\begin{figure}[!t]
\begin{center}
\includegraphics[scale=0.45]{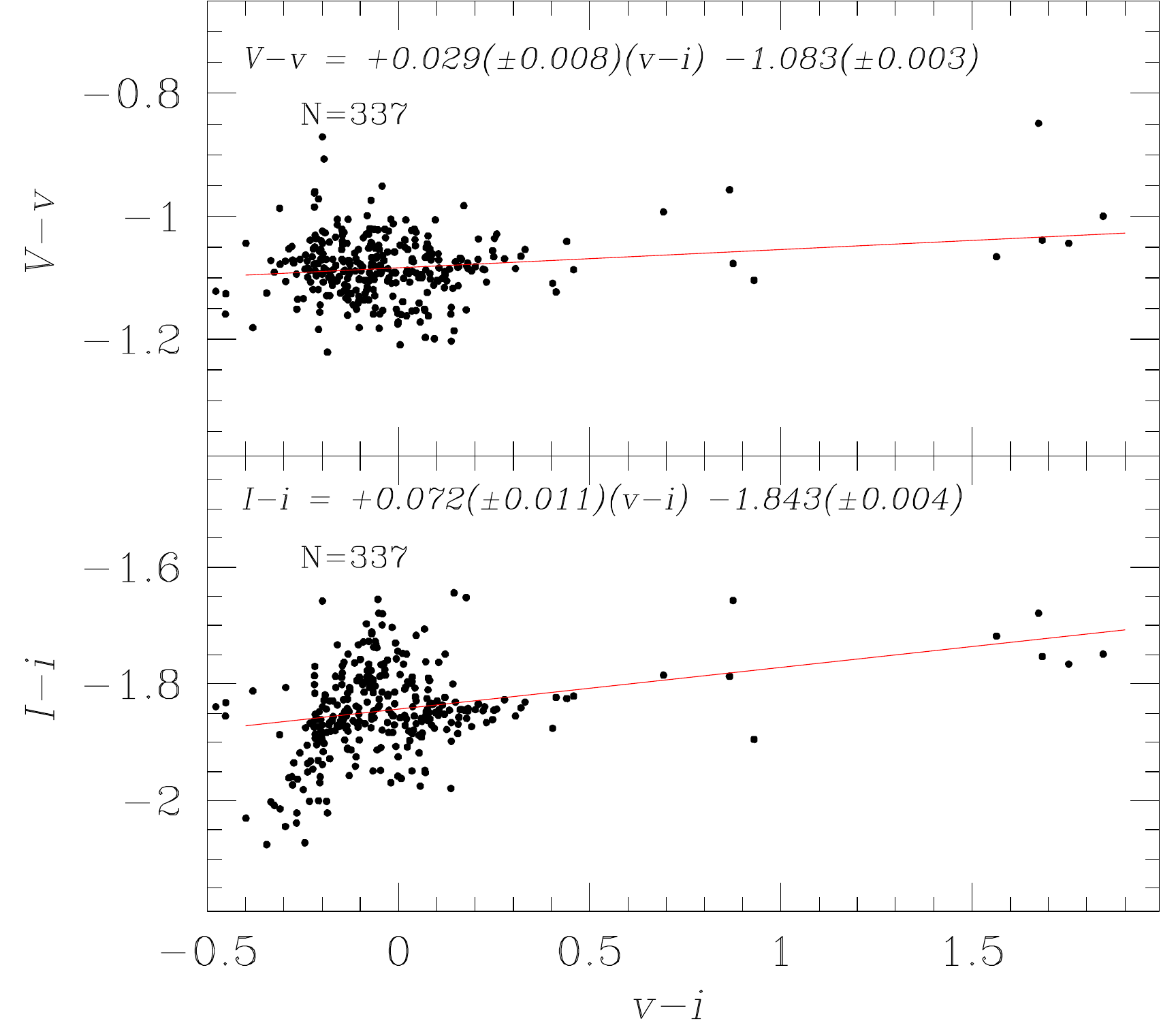}
\caption{The transformation relationship between the instrumental and standard
photometric systems using a set of standards of Stetson (2000) in the FoV of our
images of NGC~4147 .}
    \label{trans}
\end{center}
\end{figure}

\subsection{Observations}

The CCD observations in the Johnson-Kron-Cousins \emph{VI}-bands 
used in the present work, were obtained between 5 February and 29 April 2012 with 
the 2.0m-telescope at the Indian Astronomical Observatory (IAO), Hanle,
India, located at 4500~m above sea level in the Himalaya.
The detector was a Thompson
CCD of 2048$\times$2048 pixels with a scale of 0.296 arcsec/
pix, translating to a field of view (FoV) of approximately
10.1$\times$10.1 arcmin$^2$.
Our data consist of 153~$V$ and 121~$I$ images. Table
\ref{tab:observations} gives an overall summary of our observations and the seeing
conditions. 

\subsection{Difference Image Analysis}
\label{DIA}

Image data were calibrated using bias and flat-field
correction procedures. We used the Difference Image Analysis (DIA)
to extract high-precision time-series photometry of all point sources in the field of
the cluster. We used
the 
{\tt DanDIA}\footnote{{\tt DanDIA} is built from the DanIDL library of IDL routines
available at \texttt{http://www.danidl.co.uk}}
pipeline for the data reduction process (Bramich et al.\ 2013), which includes an 
algorithm that models the convolution kernel matching the PSF
of a pair of images of the same field as a discrete pixel array (Bramich 2008). 
A detailed description of the procedure is available in the paper by
 Bramich et al.\ (2011), to which the interested reader is referred for 
the relevant details.

We used the methodology of
Bramich \& Freudling (2012) to solve for the 
magnitude offset that may be introduced into the photometry 
by the error in the fitted value of the photometric scale factor
corresponding to each image.  
The magnitude offset due to this error was of the order
of $\approx 5$~mmag.

\subsection{Transformation to the \textit{VI} standard system}

From the standard stars of Stetson (2000)\footnote{%
 \texttt{http://www3.cadc-ccda.hia-iha.nrc-cnrc.gc.ca/\\
community/STETSON/standards}}
in the field of NGC~4147, we identified 337 stars in the FoV of our images, with $V$
in the range 
14.6--21.5~mag and \emph{V-I}
within $0.0$--$2.37$~mag. These stars were used to transform our instrumental system
to the Johnson-Kron-Cousins photometric system (Landolt 1992). The standard minus the
instrumental magnitude differences show a mild dependence on the colour, as displayed
in Fig.~\ref{trans}. The transformation equations are given in the
corresponding panel of Fig.~\ref{trans}. The resulting \emph{VI} photometry for all variables
in our FoV is published in Table~\ref{tab:vri_phot} which is an extract from
the full table, available in electronic format. For completenes we also include in this table
the \emph{VR} photometry from AF04, which was not made public in that publication. 

\begin{table}
\footnotesize
\caption{Time-series \textit{V}, \textit{R} and \textit{I} photometry for all the confirmed
variables in NGC 4147 taken in 2003 (AF04) and 2012 (present work). This is an extract from
the full table, which is available in electronic format.}
\centering
\begin{tabular}{cccc}
\hline
Variable &Filter & HJD & $M_{\mbox{\scriptsize std}}$ \\
Star ID  &    & (d) & (mag)      \\
\hline
 V1 & V &2452651.90900 & 16.612\\
 V1 & V &2452651.92300 & 16.613\\

\vdots   &  \vdots  & \vdots & \vdots \\
 V1 & R &2452651.912 &   16.421 \\
 V1&  R &2452651.913 &  16.439 \\

\vdots   &  \vdots  & \vdots & \vdots \\
 V1 & I& 2455963.22529 & 16.512 \\
 V1 & I& 2455963.22752 & 16.542 \\

\vdots   &  \vdots  & \vdots & \vdots \\
 V2 & V &2452651.89000 & 17.102 \\
 V2 & V &2452651.89600 & 17.139 \\

\vdots   &  \vdots  & \vdots & \vdots \\
 V2 & R &2452651.932  &  17.020 \\
 V2 & R &2452652.010  &  17.122 \\

\vdots   &  \vdots  & \vdots & \vdots \\
 V2 & I &2455963.22529 & 16.187 \\
 V2 & I &2455963.22752 & 16.197 \\
\vdots   &  \vdots  & \vdots & \vdots \\
\hline
\end{tabular}
\label{tab:vri_phot}
\end{table}

\begin{figure*}
\includegraphics[scale=0.95]{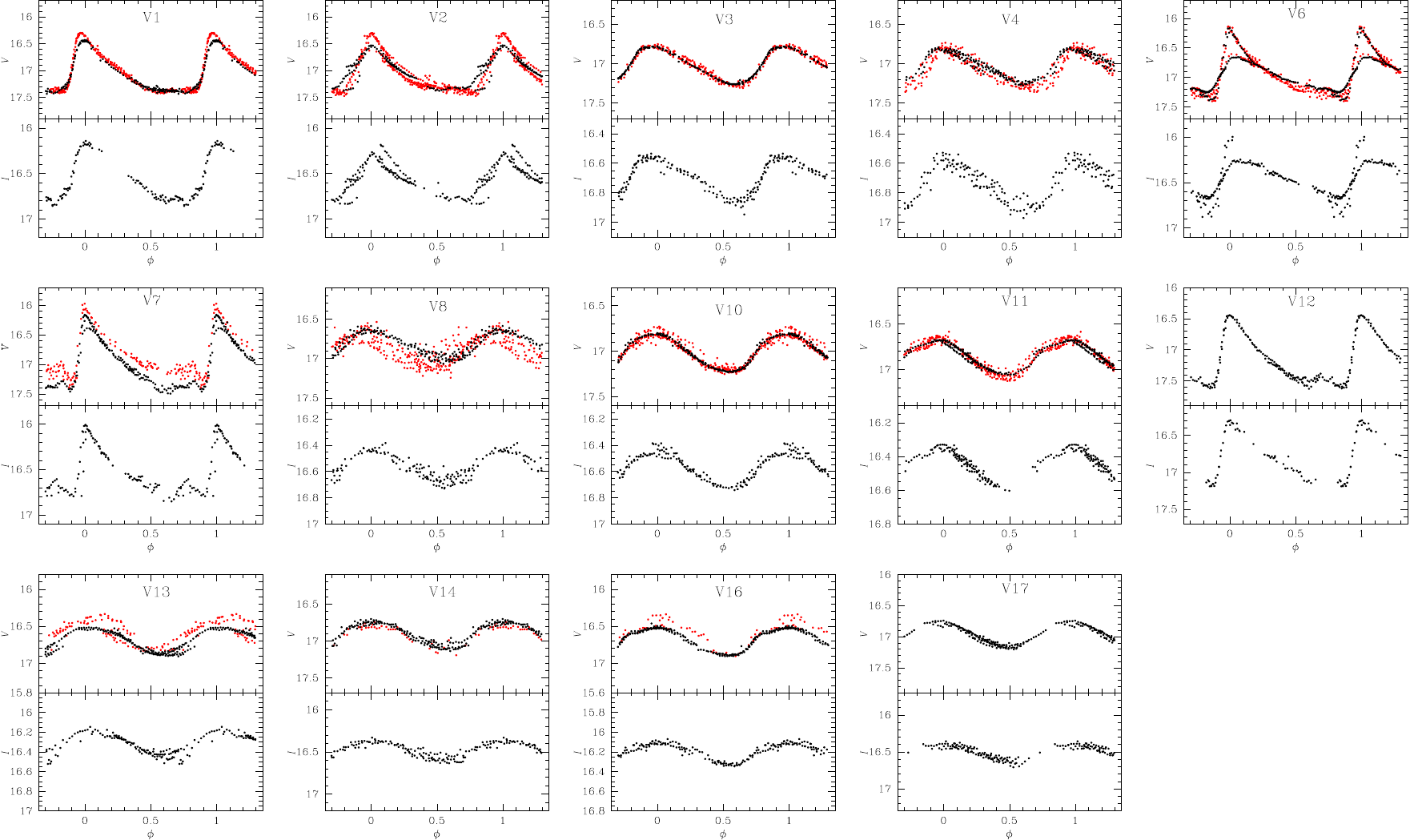}
\caption{Light curves of the RR Lyrae stars in NGC 4147
phased with the periods listed in Table~\ref{variables}. In the upper panels, black
symbols correspond to data from 2003 and red to data from 2012.}
    \label{RRL}
\end{figure*}

\section{Variable Stars in NGC 4147}

All confirmed variable stars in our FoV are listed in Table~\ref {variables}
along with their
mean magnitudes, amplitudes, and periods. The coordinates
listed in columns~10 and~11 were taken from the Catalogue of Variable Stars in
Globular Clusters (CVSGC) (Clement et al. 2001). 
For comparison we include in column 7, the periods as listed by SCS05, and it should
be noted that in some cases the periods are significantly different from the ones
found from our data, listed in column 9. For stars with light curves poorly covered
by our photometry we
have adopted the period of SCS05. The light curves of the RR Lyrae stars are shown in 
Fig. \ref{RRL}. 

\begin{table*}
\scriptsize
\begin{center}

\caption{General data for all of the confirmed variables in NGC~4147 in the \fov of 
our images.  Period estimates from SCS05 are listed in
column~7 for comparison with our periods listed in column~9.}
\label{variables}

\begin{tabular}{llllllllllll}
\hline
Variable & Variable & $<V>$ & $<I>$   & $A_V$  & $A_I$   & $P$ (SCS05) & 
HJD$_{\mathrm{max}}$ 
& $P$ (this work)    & RA   & Dec         \\
Star ID  & Type     & (mag) & (mag)   & (mag)  & (mag)   & (d)  & 
($+2\,450\,000$) & (d) & (J2000.0)   & (J2000.0)    \\
&&&&&&&&&&&\\
\hline
V1      & RRab Bl    & 17.066  &16.535  & 1.05      &0.70 & 0.500403
       & 5988.49483      & 0.500393     & 12:09:59.38 &+18:31:48.4\\   
V2      & RRab Bl    & 17.089  & 16.600  &  1.161  & 0.65   & 0.493180      
        & 5988.3470   & 0.493297      & 12:10:04.96 &+18:32:04.5\\
V3      & RRc     & 17.031  & 16.710  &  0.485  & 0.333   &0.2805427      
        & 5986.5186   &  0.280543     & 12:10:04.38 &+18:31:58.5\\
V4       & RRc     & 17.047 & 16.750  &  0.482   & 0.389   & 0.300031      
        & 5963.2723    & 0.300066      & 12:10:06.39 &+18:32:50.2\\
V6       & RRab Bl      & 16.941 &16.455  &1.208   & 0.810 & 0.609730    
        & 5963.3748       & 0.609737      & 12:10:08.51 &+18:33:00.0\\
V7       & RRab      & 17.066 & 16.567 &1.301    & 0.823  & 0.514245       
        & 5986.4396     & 0.514321   & 12:10:06.66 &+18:32:39.7\\    
V8       & RRc    &16.828   & 16.558  & 0.431 & 0.309   & 0.278652  
       & 5963.4587     & 0.278599   & 12:10:06.95 &+18:32:34.8\\
V10      & RRc      &17.020 & 16.585  & 0.421  & 0.302 & 0.352301      
       & 5963.2378      & 0.352339    & 12:10:03.74 &+18:31:48.4\\
V11       & RRc     & 16.867 &16.471 &  0.380  & 0.269   & 0.387419        
        & 5963.3836    &0.387423  & 12:10:05.53 &+18:31:51.8\\
V12       & RRab     &17.199 & 16.858  & 1.164   & 0.882  & 0.504700     
        & 5986.5132     &0.504700  & 12:10:06.70 &+18:32:28.4\\  
V13       & RRc    & 16.701  &16.322  & 0.385  & 0.282  & 0.408320    
       & 5963.3660      & 0.408319  & 12:10:06.37 &+18:32:14.1\\
V14       & RRc      & 16.914  &16.483 &0.432  & 0.233  & 0.356376       
       & 5963.4586   & 0.356375  & 12:10:06.94 &+18:32:32.4\\  
V16       & RRc      & 16.707  &16.211 &0.388  & 0.257  & 0.372259        
       & 5963.4587    & 0.372134    & 12:10:07.35 &+18:32:40.1\\
V17      & RRc    & 16.968 &16.512 & 0.453  & 0.285   &  0.371229 
       & 5988.3095       & 0.374843 & 12:10:10.67 &+18:34:51.2\\
V18       & SR    & 13.929$^a$ & 12.391$^a$ & 0.2   & --   &--     
       & --     & 24.8  & 12:10:05.63 &+18:32:11.6\\
V19       & RRc    & 17.03$^a$ &16.72$^a$ &0.28$^a$ & 0.14$^a$& 0.273933     
       & --       & --   & 12:10:21.98 &+18:35:02.1\\
\hline
\end{tabular}
\raggedright
\center{\quad \emph{a}: Value taken from SCS05.}
\end{center}
\end{table*}

\begin{figure}
\begin{center}
\includegraphics[scale=2.0]{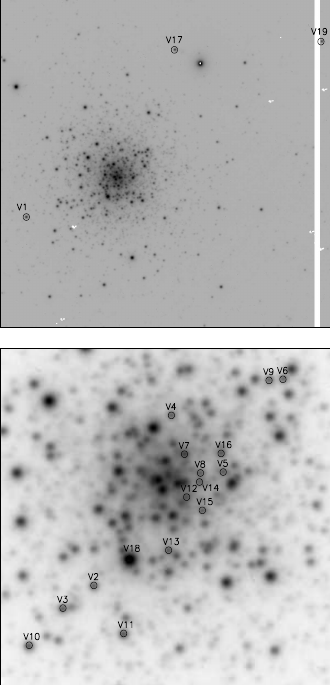}
\caption{Identification charts of all known variables in the field of NGC 4147. The
field size in the upper and lower frames is 6$\times$6 and 1.5$\times$1.5 arcmin$^2$
respectively. In both frames North is up and East is to the right.}
    \label{chart}
\end{center}
\end{figure}

All known variables in NGC 4147 are identified in the charts of Fig. \ref{chart} for
the cluster peripheral and core regions.

\subsection{Reddening}
\label{redd}

It is well known from the work of Sturch (1966) that RRab stars have nearly 
the same intrinsic colour$(B-V)_0$ at minimum light. The value $(V-I)_{0;min} =
0.58\pm0.02$ was calibrated by Guldenschuh et al. (2005). We have used these results
to calculate the reddening of four RRab stars in NGC 4147 from their minimum
value in \emph{V-I} and found an average $E(V-I) = 0.010 \pm 0.003$. Taking 
\emph{E(V-I)/E(B-V)}= 1.259 (Schlegel et al. 1998) we get an average 
\emph{E(B-V)}$= 0.008 \pm 0.002$. This value is a little smaller than the value 0.02,
listed by Harris (1996) or the value given by the reddening maps of Schlafly \&
Finkbeiner (2011) 0.022.
We shall adopt \emph{E(B-V)}= 0.01 in the remainder of the paper.

\section{Physical parameters of RR~Lyrae stars}
\label{sec:Four}

Physical quantities of individual RR Lyrae stars can be estimated via the Fourier 
decomposition of the light curve and the employment of ad-hoc well-tested
calibrations.
A full description of the method and the specific calibrations and their zero points 
used in a homogeneous way 
in several of our previous papers for a family of globular clusters, can be seen
in the paper by Arellano Ferro, Bramich \& Giridhar (2017). For briefness we do not
reproduce the
details here, however, for the sake of clearness in the nomenclature we remind that
the light
curve can be represented by the equation,

\begin{equation}
\label{eq.Foufit}
m(t) = A_0 + \sum_{k=1}^{N}{A_k \cos\ ({2\pi \over P}~k~(t-E) + \phi_k) },
\end{equation}
%
%\noindent
where $m(t)$ is the magnitude at time $t$, $P$ is the period, and $E$ is the epoch. A
linear
minimization routine is used to derive the best fit values of the 
amplitudes $A_k$ and phases $\phi_k$ of the sinusoidal components. 
From the amplitudes and phases of the harmonics in Eq.~\ref{eq.Foufit}, the 
Fourier parameters, defined as $\phi_{ij} = j\phi_{i} - i\phi_{j}$, and $R_{ij} =
A_{i}/A_{j}$, 
are computed. 

We have argued in previous papers in favour of the calibrations developed by Jurcsik
\& Kov\'acs (1996) and Kov\'acs \& Walker (2001) for the iron abundance and absolute
magnitude of RRab stars, and those of Morgan, Wahl \& Wieckhorts (2007) and Kov\'acs
(1998) for RRc stars. 
The effective temperature $T_{\rm eff}$ is estimated using the calibration of Jurcsik
(1998). These calibrations and their zero points have been discussed in detail in
Arellano Ferro et al.\ (2013).

The calculation of the physical quantities is very sensitive to the morphology of the
light curve
and spurious results can be obtained if light curves with peculiarities are
considered, for example
those with amplitude and/or phase modulations, or cases of incomplete
light curves.
In particular, for the calculation of [Fe/H] in RRab stars, Jurcsik \& Kov\'acs (1996)
and  Kov\'acs \& Kanbur (1998) have defined a "compatibility parameter" $D_m$ which
should be smaller than 3.0 
if the light curve under analysis is consistent with the calibrators used to establish
the
 Jurcsik \& Kov\'acs (1996) iron abundance calibration.
In the present case we note that all five known RRab star show amplitude variations and
their $D_m$
values are all very large (column 10 in Table \ref{tab:fourier_coeffs}), hence they
are neither fit for 
[Fe/H] calculation nor for a distance estimation since the absolute magnitude depends
on 
the amplitudes of the first and third harmonics (Kov\'acs \& Walker 2001), thus they
were ignored. Likewise, some peculiar RRc stars with erratic phasing (V8, V13 and V16)
were also not included
in the physical parameters calculations. 
 
Nevertheless, the value of $A_0$ and the Fourier light-curve fitting parameters
for the 5 RRab and 9 RRc stars are given in Table~\ref{tab:fourier_coeffs}. When
amplitude 
variations are present, we have taken care in fitting the maximum amplitude curve. The
corresponding 
physical parameters for the most stable RRc stars are listed in 
Table~\ref{fisicos}. The absolute magnitude $M_V$ was converted into luminosity with
$\log (L/{\rm L_{\odot}})=-0.4\, (M_V-M^\odot_{\rm bol}+BC$). The bolometric
correction was
calculated using the formula $BC= 0.06\, {\rm [Fe/H]}_{ZW}+0.06$ given by Sandage \&
Cacciari (1990). We adopted  $M^\odot_{\rm bol}=4.75$~mag. For the distance
calculation, we have adopted \emph{E(B-V)}=0.01, estimated in $\S$ \ref{redd}.

\begin{table*}
\scriptsize
\begin{center}
\caption[] {Fourier coefficients of  \RRab and \RRc stars in NGC~4147. 
The numbers in parentheses indicate
the uncertainty on the last decimal place. Also listed is the
deviation parameter~$D_{\textit{\lowercase{m}}}$ for the \RRab stars.}       
\label{tab:fourier_coeffs}   
\begin{tabular}{lccccccccc}
\hline
Variable     & $A_{0}$    & $A_{1}$   & $A_{2}$   & $A_{3}$   & $A_{4}$   &$\phi_{21}$
& $\phi_{31}$ & $\phi_{41}$ 
& $D_m$ \\
  ID     & ($V$ mag)  & ($V$ mag)  &  ($V$ mag) & ($V$ mag)& ($V$ mag) & &  & & \\
\hline
%\multicolumn{1}{c}{RRab} \\
&&&&&RRab&&&&\\
\hline
V1 &17.066(2)& 0.440(3)& 0.164(3)& 0.131(3)&0.084(3)&
3.855(21)&7.981(31)&5.864(45)&16.7\\
V2 &17.089(4)& 0.413(5)& 0.229(6)& 0.108(6)&0.054(6)&
3.674(40)&7.581(73)&5.435(126&8.4\\
V6 &16.941(2)& 0.343(4)& 0.211(4)& 0.152(4)&0.105(4)&
4.092(30)&8.314(45)&6.241(62)&22.7\\
V7 &17.066(5)& 0.454(7)& 0.177(7)& 0.145(7)&0.108(7)&
3.850(51)&7.722(69)&5.686(92)&12.4\\
V12&17.197(3)& 0.431(4)& 0.195(4)& 0.146(4)&0.088(4)&
3.781(29)&7.933(41)&6.071(61)&20.5\\
\hline
%\multicolumn{1}{c}{RRc} \\
&&&&&RRc&&&&\\
\hline

V3& 17.031(1)& 0.229 (2)& 0.051 (2)& 0.012 (2)& 0.009 (1)& 4.567  (49)& 2.470 (188)&
0.942 (262)&\\
V4& 17.047(2)& 0.202 (3)& 0.053 (3)& 0.016 (3)& 0.019 (3)& 4.478  (78)& 1.979 (242)&
0.413 (209)&\\
V8& 16.828(3)& 0.178 (4)& 0.032 (4)& 0.008 (4)& 0.007 (4)& 4.561 (157)& 2.599 (606)&
1.342 (680)&\\
V10&17.020(1)& 0.211 (1)& 0.025 (1)& 0.014 (1)& 0.009 (1)& 4.672  (63)& 3.616 (118)&
2.124 (181)&\\
V11&16.867(1)& 0.186 (1)& 0.007 (1)& 0.015 (1)& 0.007 (1)& 5.252 (193)& 4.243  (95)&
2.242 (185)&\\
V13&16.701(1)& 0.181 (2)& 0.004 (2)& 0.013 (2)& 0.004 (2)& 5.833 (701)& 2.053 (204)&
5.861 (605)&\\
V14&16.914(2)& 0.176 (3)& 0.021 (3)& 0.016 (3)& 0.005 (3)& 4.718 (173)& 3.879 (230)&
1.988 (732)&\\
V16&16.707(1)& 0.192 (1)& 0.014 (1)& 0.018 (1)& 0.011 (1)& 5.134 (120)& 3.902  (99)&
2.713 (159)&\\
V17&16.968(2)& 0.198 (3)& 0.013 (3)& 0.006 (3)& 0.006 (3)& 4.728 (311)& 3.490 (712)&
5.884 (589)&\\

\hline
\end{tabular}
\end{center}
\end{table*}

\begin{table*}
\footnotesize
\begin{center}
\caption[] {\small Physical parameters of the \RRc stars. The
numbers in parentheses indicate the uncertainty on the last 
decimal places.}
\label{fisicos}
 \begin{tabular}{lcccccccc}
\hline 
Star&[Fe/H]$_{ZW}$ & [Fe/H]$_{UVES}$ &$M_V$ & log~$T_{\rm eff}$  &log$(L/{\rm
L_{\odot}})$ &$D$ (kpc)& 
$M/{\rm M_{\odot}}$&$R/{\rm R_{\odot}}$\\
\hline
V3&-1.57(31)&-1.49(35)&0.620(5)&3.869(1)&1.652(2)& 18.88(4)& 0.59(1)& 4.12(2)\\
V4&-1.91(41)&-1.96(56)&0.560(14)&3.862(1)&1.676(6)& 19.55(13)& 0.62(1)& 4.37(3)\\
V10&-1.74(24)&-1.72(30)&0.546(5)&3.861(1)&1.682(2)& 19.44(5)& 0.51(1)& 4.42(1)\\
V11&-1.72(21)&-1.70(26)&0.496(10)&3.858(1)&1.702(4)& 18.54(8)& 0.48(1)& 4.58(2)\\
V14&-1.65(47)&-1.60(56)&0.558(15)&3.861(1)&1.677(6)& 18.41(13)& 0.49(1)& 4.38(3)\\
V17&-1.95(33)&-2.01(36)&0.535(19)&3.856(4)&1.686(8)& 19.07(17) &0.49(3)& 4.54(4)\\
\hline
Weighted mean&
$-$1.72(13)&$-$1.68(16)&0.571(3)&3.861(1)&1.671(1)&19.05(3)&0.52(1)&4.30(1)\\
$\sigma$&$\pm$0.15&$\pm$0.15&$\pm$0.040&$\pm$0.004&$\pm$0.016&$\pm$0.46&$\pm$0.06&$\pm
$0.16\\
\hline
\end{tabular}
\end{center}
\end{table*}

The resulting physical parameters of the RRc stars are
summarized in Table~\ref{fisicos}. The mean values given in the bottom of the table
are weighted by the statistical uncertainties. The iron abundance is given in the
scale of Zinn \& West (1984) and in the scale of Carretta et al. (2009). The
transformation between these two scales is of the form:

\begin{eqnarray}\label{UVES}
[\mathrm{Fe/H}]_{UVES} &=& -0.413 +0.130\, [\mathrm{Fe/H}]_{ZW} \nonumber \\
  && -0.356\, [\mathrm{Fe/H}]_{ZW}^2.
\end{eqnarray} 

Also listed are the corresponding distances.
Given the period, luminosity, and temperature for each RR~Lyrae star, its
mass and radius can be estimated from the equations: $\log~M/M_{\odot} =
16.907 - 1.47~ \log~P_F + 1.24~\log~(L/L_{\odot}) - 5.12~\log~T_{\rm eff}$ (van Albada
\& Baker 1971), and $L$=$4\pi R^2 \sigma T^4$ respectively.
The masses and radii given in Table~\ref{fisicos} are expressed in solar
units.

\begin{figure*}
\begin{center}
\includegraphics[scale=1.5]{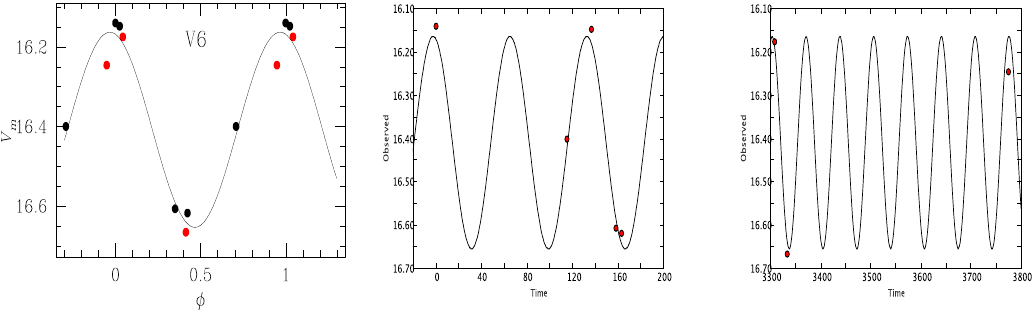}
\caption{The left panel shows the amplitude modulation of the $V_m$ magnitude in V6
phased with a period of 67.46d; black symbols are
observations from 2003 (AF04) and red symbols are 2012 observations from the present
work. In the central and right panels the time scale in the horizontal axis is
HAD$-$2452654.935 and the sinusoidal curve has a period of 67.46d.}
\label{TmaxV6}
\end{center}
\end{figure*}

\section{Notes on individual stars.}
\label{indstars}

In the following paragraphs we comment on certain peculiarities of specific anomalous
stars.

V6. This RRab star has pronounced Blazhko amplitude modulations. Analysing
the magnitude at maximum light $V_m$ variations, AF04 estimated a likely Blazhko
period of about 67.9d. In the
present data we have identified three more times of maximum. In Table \ref{Tab:TmaxV6}
we list the times of maximum and their corresponding magnitudes $V_m$ . The
$V_m$ modulations are shown in first panel of Fig. \ref{TmaxV6}. We calculated a
periodicity of 67.46d which agrees very well with the original estimation of AF04.
In the central and right panels one can see that the
sinusoidal model with $P=67.46$d does fit the $V_m$  modulations very well.

V12. In the paper of AF04 the $V$ light curve is rather incomplete but shows a very
large amplitude, 
likely due to the presence of four spurious data near minimum light.
The light curve of SCS05 is of smaller amplitude with a flat bottom, attributed by
SCS05 to
the possible presence of an unseen companion. In the present paper the $V$ light curve
(Fig. \ref{RRL})
is clean and complete, with an amplitude similar to that of SCS05. A
comparison of 
all available data suggest some small amplitude modulations. The intensity weighted
colour $<V>$-$<I>$ from our data is 0.340 and in the CMD (Fig. \ref{CMD}) the
star falls in the inter-mode region. The colour from SCS05 is 0.48 which moves the
star to
the fundamental mode region. Given the uncertainties in the colour estimates
of this peculiar star, we chose to omit it in the discussion involving the RRab-RRc
distribution in the instability strip.

\begin{figure}
\begin{center}
\includegraphics[scale=0.4]{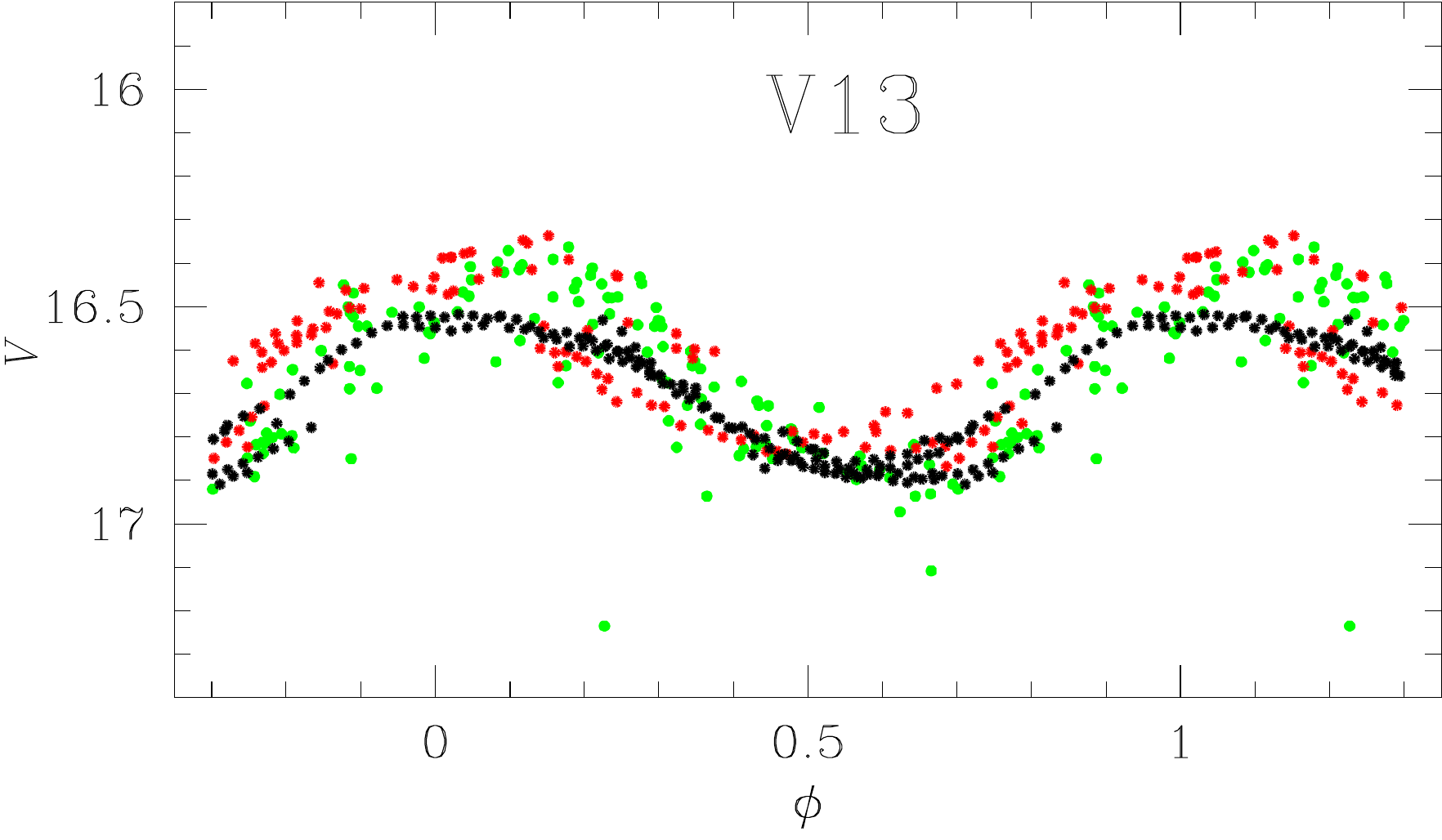}
\caption{$V$ light curve of V13. Green symbols are data from SCS05, red are from
AF04 and black are data from the present paper.}
    \label{V13all}
\end{center}
\end{figure}

V13. This RRc star was noted by AF04 to show amplitude modulations. SCS05 in their
investigation calculated a period of 0.408319d but comment that the resulting
light curve is somewhat noisy. Combining AF04, SCS05 and the present paper data we
found a period of 0.408318d, in excellent agreement with that of SCS05. The resulting
phased $V$ and $I$ light curves are shown in Figs. \ref{RRL} and \ref{V13all}.
In the later we include the $V$ data from SCS05. The $V$ light curve displays
amplitude
and phase displacements. It has been suggested by SCS05 that these erratic phasing may
be the consequence of interference of fundamental and first overtone pulsation modes
but they failed identifying the two modes. An alternative explanation might be the
interference of two modes of very similar frequencies, and at least one of them being
non-radial, similar to the case of V37 in NGC 6362 (Smolec et al. 2017;  Arellano
Ferro et al. 2018). Detecting the
two frequencies requires dense high quality observations that are not presently
available for V13. Possible light contamination by an unseen companion also cannot be
ruled out.

V16. Similarly to V13, this star shows erratic phasing, which was also noted by SCS05.
It falls about 0.02 mag to the red of the red edge of the first overtone instability
strip (Fig. \ref{CMD}), but its position may be spurious given the amplitude and
phase variations.

% TABLA 3
\begin{table}
\scriptsize
\begin{center}

\caption{Variation of the magnitude at maximum $V_{m}$ in V6 due to
the Blazhko effect (see Fig. 3).}
\label{Tab:TmaxV6}

\begin{tabular}{cc}
\hline
HJD$_{max}$ & $V$ \\
\hline
2452654.935    & 16.140\\
2452770.192    & 16.400\\
2452791.514    & 16.148\\
2452813.481    & 16.607\\
2452818.382    & 16.618\\
2455963.366    & 16.175\\
2455988.398    & 16.666\\
2456029.223    & 16.246\\
\hline
\end{tabular}
\end{center}
\end{table}

\begin{figure}
\begin{center}
\includegraphics[scale=1.2]{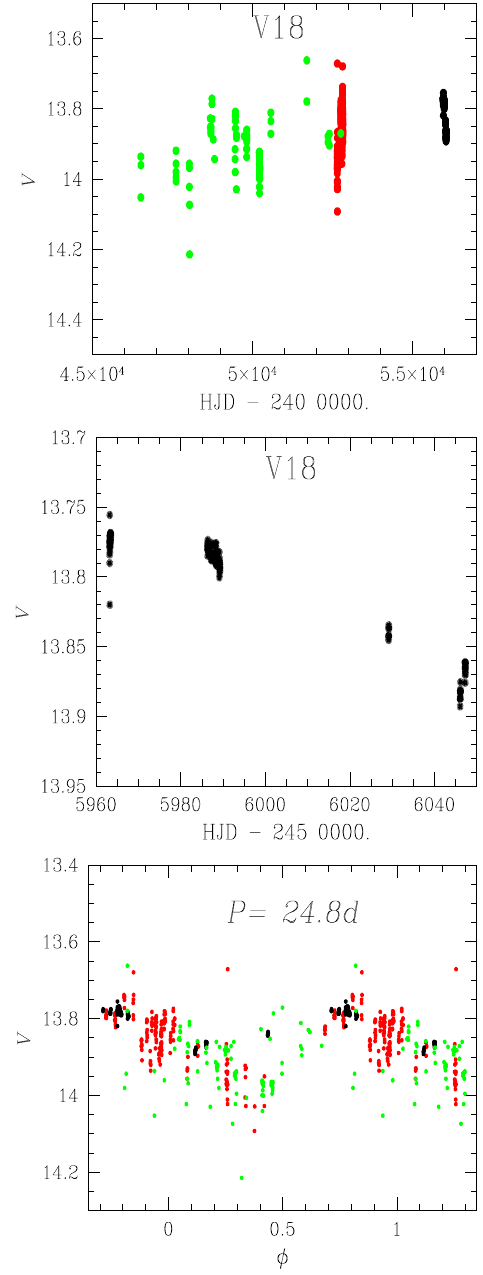}
\caption{Historical light variations in the $V$-band of the star V18. Top panel
shows a long term brightening over the last nearly 27 years. Green circles are data
from SCS05, red from AF04 and
black from the present paper. The middle panel is an enlargement of the 2012 data that
clearly show a slow light dimming. This demonstrates the variable nature of the star.
The bottom panel shows the light curve with phase with a period of 24.8d. We
classify the star as semi-regular red variable of the SR type.}
    \label{V18}
\end{center}
\end{figure}

V18. This is the brightest star in the field of the cluster near the central region.
 The star was classified by AF04 as a
possible RRc and later denied as variable by SCS05, who considered the star a bright
red giant but failed finding a convincing variability. We have reconsidered the case
employing the historical data of SCS05 taken between 1983 and 2003 by an assortion of
instruments, the data from AF04 taken during 24 nights between January and July,
2003, and the data of the present paper taken in 2012 as detailed in Table
\ref{tab:observations}. In Fig. \ref{V18}, the top panel shows the
long-term brightening of the star, while the panel in the middle shows the clear
dimming in 2012. This demonstrates beyond doubt the variable nature of the star. The
following questions are if the star is periodic and what the period might be. We have
performed a series of test for several data blocks and found that the AF04 and 2012
data are well phased by a period of about 46.1d, the combination of all available $V$
data suggests about half that period, 24.8d. The bottom panel shows the light curve
phased with this period. 

The star is saturated in our $I$-band images, hence we cannot calculate \emph{V-I}. 
Adopting the $<V>$ and $<I>$ from SCS05 we were able to plot the star on the CMD
of
Fig. \ref{CMD}. The star is near the tip of the red giant branch (RGB) as pointed out
by SCS05. Based on these evidences we classify the star as a semi-regular late-type
variable (SR) whose characteristic time of variation or semi-period is to be
better estimated when new accurate data become available.

V19. This RRc star discovered by SCS04 is near the edge of our images and very near
to a bad pixel column of our CCD chip, thus we were unable to secure any confident
photometry.

\section{Distance to NGC 4147 from its variable stars}
\label{sec:DISTANCE}

The distance to NGC 4147 obtained from the Fourier decomposition of RRc stars is 
19.05$\pm$0.46 kpc.

Also for the RR Lyrae stars one can make use of the P-L relation for the $I$
magnitude derived by Catelan et al. (2004):

 \begin{equation}
M_I = 0.471 - 1.132~{\rm log} P + 0.205~{\rm log} Z,
\label{eqn:PL_RRI}
\end{equation}

\noindent
with ${\rm log}~Z =[M/H] -1.765$ and $[M/H] = \rm{[Fe/H]} - \rm {log} (0.638~f +
0.362)$ and log~f = [$\alpha$/Fe] (Salaris et al. 1993). For the sake of direct
comparison with the results of
Fourier decomposition we applied the above equations to the $I$ measurements of the 6
RRc stars in 
Table \ref{fisicos} and found  a mean distance of 18.89$\pm$0.57 kpc, in good
agreement with the Fourier results.

In principle, another approach to estimate the cluster distance is using the variables
near the tip of the RGB (TRGB). We can take advantage of the fact that V18 is
semi-regular
red variable or SR type, and sits near the TRGB.
The method, originally developed to estimate distances to nearby galaxies (Lee et
al. 1993) has already
been applied by our group for the distance estimates of other clusters e.g. Arellano
Ferro et al. (2015) for
NGC 6229 and Arellano Ferro et al. (2016) for M5 and it is described in detail in
these papers. 
In brief, the idea is to use the bolometric magnitude
of the tip of the RGB as an indicator, which can be estimated using the calibration of
Salaris \& Cassisi
(1997):

\begin{equation}
\label{TRGB}
M_{bol}^{tip} = -3.949\, -0.178\, [M/H] + 0.008\, [M/H]^2,
\end{equation}
 
\noindent
where $[M/H] = \rm{[Fe/H]} - \rm {log} (0.638~f + 0.362)$ and log~f = [$\alpha$/Fe] 
(Salaris et al. 1993). However, as argued by Viaux et al. (2013)
in their analysis of M5, the brightest stars on the
RGB might not be exactly at the TRGB but some dimmer leading to an overestimation of
the distance.
According to these
authors the TRGB may be between 0.05 and 0.16 mag brighter than the brightest stars on
the
RGB. For V18 in NGC 4147, if we assume that it is exactly at the TRGB, which is
unlikely,
we find a distance of 20.3 kpc. To bring this distance into agreement with the
results from the RRc star we must assume that V18 is 0.14 mag below the true TRGB, in
this case we find a distance of 19.08 kpc. 

The distances found above are in excellent agreement with the values listed
by Harris (1996) of 19.3 kpc and by SCS05 of 19.05 kpc.

\section{The CMD of NGC~4147}
\label{sec:CMD}

\begin{figure*}
\begin{center}
\includegraphics[scale=0.7]{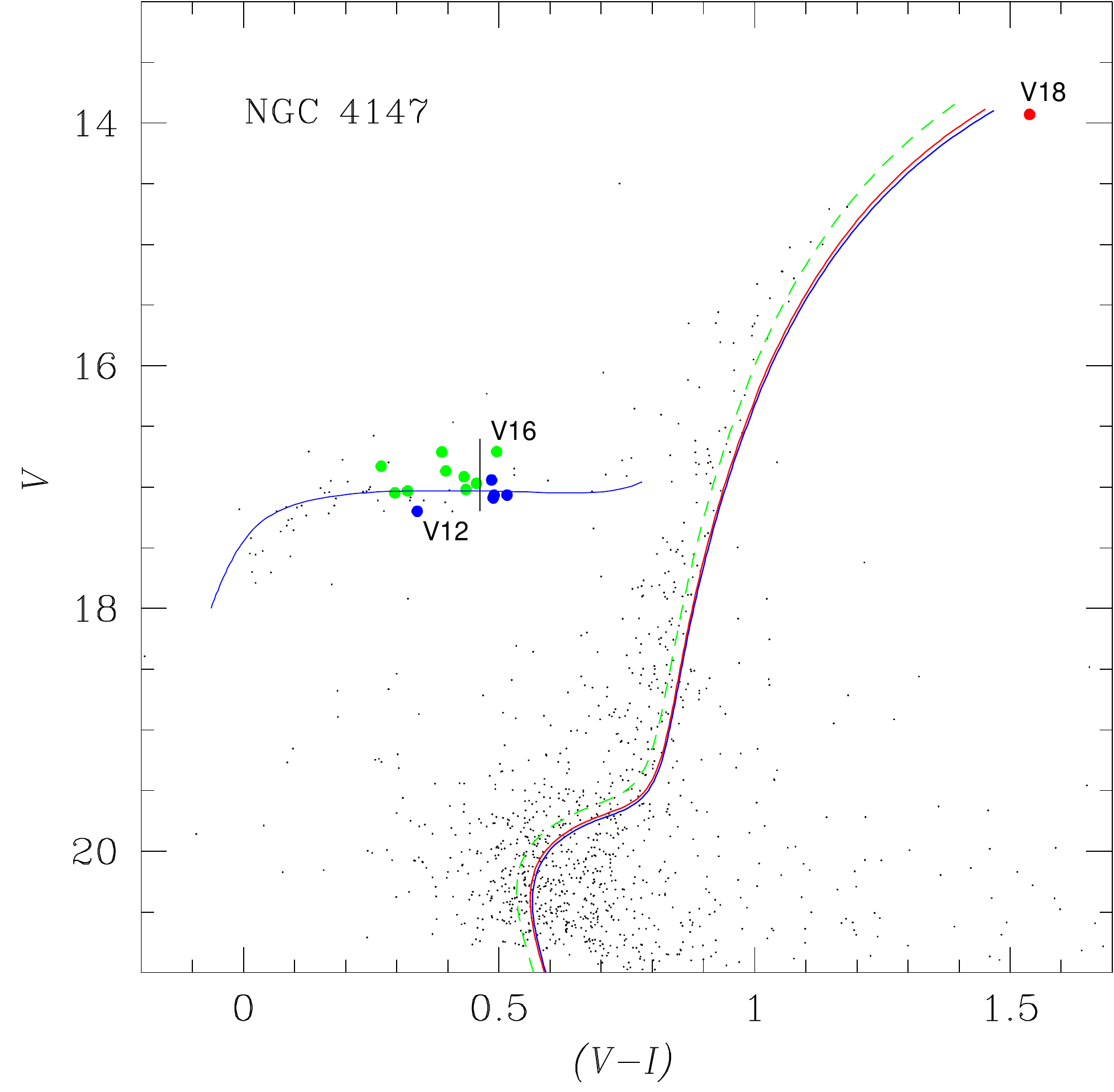}
\caption{CMD of NGC 4147. The plotted magnitudes and colors for all
non-variable stars are magnitude-weighted means of their light curves. Variables are
plotted using their intensity-weighted means $<V>$ and $<V>-<I>$. Symbols and
colors are: blue and green circles RRab and RRc stars respectively. The SR star
V18 is shown with a red symbol. Since the star is saturated in our $I$ images we used
the values of SCS05 to plot the star. Isochrones for 12 Gyr,
Y=0.25 and [$\alpha$/Fe]=0.4 for metallicities --1.75 (blue) and --1.80 (red),
and the ZAHB  for [Fe/H]=--1.75, shifted to a distance of 19.05 kpc and
$E(B-V)=0.01$ come from the model collection of VandenBerg et al. (2014). For
comparison the green segmented isochrone is for [Fe/H]=--2.0 and assuming 
$E(B-V)=0.0$. 
The black vertical line in the HB is the empirical red edge of the first overtone
instability strip at $(V-I)_0
\sim 0.45$ (Arellano Ferro et al. 2016) duly reddened. See text for a discussion.}
\label{CMD}
\end{center}
\end{figure*}

The CMD of the cluster is shown in Fig.~\ref{CMD}, where the location of all 
known variables is marked. All variable stars are plotted using their
intensity-weighted means $< V >$ and corresponding colour $<V>-<I>$. 

The age of NGC 4147 has been estimated via a differential approach by De Angeli et al.
(2005) as 12.23 Gyrs. On the other hand, SCS05 have argued that the ages of M3 and NGC
4147 are virtually identical. For M3 De Angeli et al. (2005) estimated 12.0 Gyrs.
We have adopted 12.0 Gyrs with the aim of overplotting an isochrone on our
CMD.

We have plotted the isochrone and ZAHB models
from VandenBerg et al. (2014) for [Fe/H]=--1.75, Y=0.25 and [$\alpha$/Fe]=0.4.
The models were shifted according to the distance and reddening 
calculated in previous sections from the RRc and RRab stars respectively,
i.e. the model parameters closely correspond to our determinations [Fe/H]=$-1.72$,
19.05 kpc (apparent
distance modulus of 16.43) and $E(B-V)$=0.01. Immediate peculiarities stand out
from this exercises: The isochrone does not fit the RGB but falls to the red by about
0.05 mag (blue isochrone in Fig.~\ref{CMD}). Note that assuming $E(B-V)$=0.02
would make things worse. Adopting [Fe/H]=--1.80 from the listing of Harris (1996, 2010
edition) the situation improves only a little (red isochrone).
It would be necessary, with this distance and reddening, to claim a much 
lower [Fe/H] to better fit the RGB. For the sake of comparison we plotted
the
isochrone for  [Fe/H]=--2.0 and $E(B-V)$=0.0 (green dashed isochrone), while it gives
better fit at RGB, it is at odds with our determined parameters.

If we look at the ZAHB, the peculiarities continue, the blue tail of the HB
bents down at a redder colour than the theoretical model by about 0.1 mag. We stress
that
the RR Lyrae positions are correct since the red edge of the first overtone mode,
i.e. the vertical black line in the CMD, falls exactly where it has been 
observed in other clusters. If we adopt $E(B-V)$ of 0.02 the problem is lessened (not
fully solved) at the expense of a larger red wards shift of the isochrone

We have not found a $(V-I)$ CMD for NGC 4147 in the literature for a direct
comparison.
A comparison with the $(B-V)$ and $(B-I)$ CMD's from SCS05 is not illustrative since
their isochrone and ZAHB are fiducial hand draws and are not suitable for comparison
with theoretical models.

\section{Oosterhoff type and the HB of NGC~4147}
\label{sec:HB}

\subsection{The Oosterhoff type}
\label{Ootype}

\begin{figure}
\begin{center}
\includegraphics[scale=0.68]{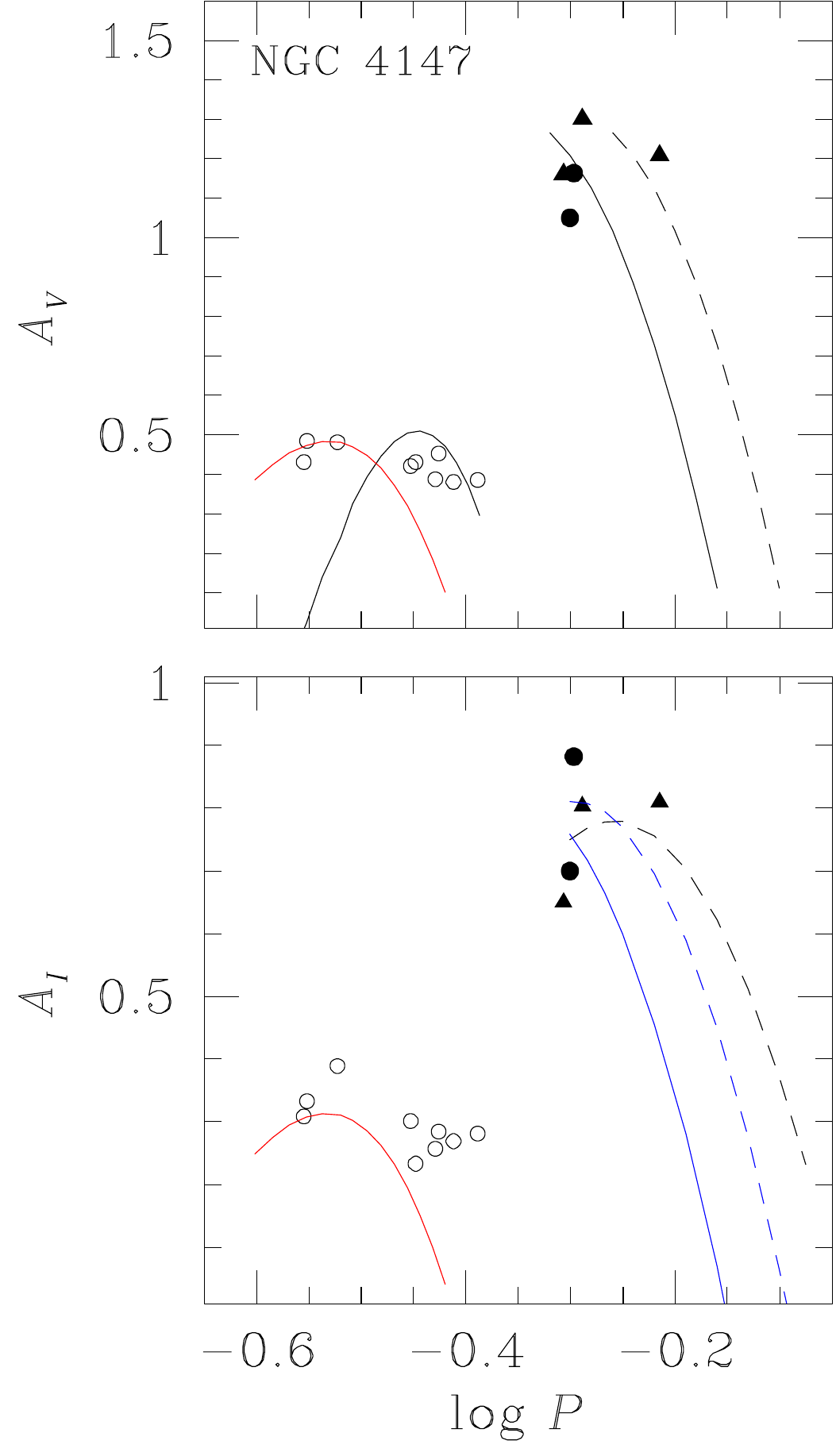}
\caption{The Period-Amplitude plane in the \emph{VI}-bands for NGC 4147.
Filled and open symbols represent RRab and
RRc stars, respectively. Triangles represent stars with
Blazhko modulations. In the top panel the continuous
and segmented lines are the loci for evolved and unevolved
stars in M3 according to Cacciari et al. (2005).
The black parabola was found by Kunder et al. (2013a)
from 14 OoII clusters. The red parabolas were calculated
by Arellano Ferro et al. (2015) from a sample of
RRc stars in five OoI clusters and avoiding Blazhko variables.
In the bottom panel the black segmented locus
was found by Arellano Ferro et al. (2011; 2013) for the
OoII clusters NGC 5024 and NGC 6333. The blue loci
are from Kunder et al. (2013b).}
\label{Bailey}
\end{center}
\end{figure}

NGC~4147 has been considered as an Oo~I type cluster in spite its low metallicity and
the rather large fraction of RRc stars, $N_c/(N_c + N_{ab}) = 0.67$, typical of Oo~II
clusters. The reason for this is the low average period of its RRab stars, $<P_{ab}>
= 0.524 \pm 0.048$~d, which unfortunately is based only on its five known RRab stars.
The period-amplitude plane, or Bailey diagram, for NGC~4147 is shown in Fig.
\ref{Bailey} in the \emph{VI}-bands. The details are given in the caption. Given the
low number of RRab stars and their strong amplitude modulations, it is not possible
to decide from their distribution, whether they are confined preferentially towards
the Oo~I locus defined from the M3 globular cluster (Cacciari et al. 2005). The
distribution of RRc stars is more suggestive and they display a dual distribution
about the loci for Oo~I (red parabola) and Oo~II (black parabola). This reminds the
distribution for NGC 6402 (Contreras Pe\~na et al. 2018; their Figures 9 and 10) for
which arguments in favour of being an Oo-Int cluster were offered.

Another interesting feature of NGC 4147 is its HB structure parametrised by the
quantity $\mathcal L$ = (B - R)/(B + V + R) with B, V, R, 
representing the number of stars on the HB to the blue, inside and to the red of the
instability strip. Fig. \ref{CATELAN} displays the positions of Galactic
globular clusters on the $\mathcal L$-[Fe/H] plane. The different Oosterhoff types
are indicated. Similar versions of this figure have been used to illustrate the
Oosterhoff dichotomy in Galactic globular clusters which is not present in
extragalactic clusters and spheroidal galaxies (e.g. Bono et al. 1994, Catelan
2009,
Contreras Pe\~na 2018). The upper and lower limits of the Oostherhoff gap are shown by
the solid lines in the figure calculated by Bono et al. (1994). This region is
mostly
populated by extragalactic systems, for instance those listed in Table 3 of Catelan
(2009). For the sake of clearness we only display Galactic globulars in Fig.
\ref{CATELAN}.

The position of NGC~4147 inside the Oosterhoff gap, although marginal, is remarkable
and consistent with
the classification of the cluster as an Oo-Int type from the RRc stars amplitude
distributions. This together with association of the cluster to the Sagittarius
dSph galaxy stream discussed by Bellazzini et al. (2003a, 2003b), supports the
scenario that NGC~4147 was accreted by our Galaxy during the merging of the
satellite Sagittarius dSph galaxy.

From, Fig. \ref{CATELAN}, three more clusters in the Oosterhoff gap call our
attention;
M14 which has been recently argued to be an Oo-Int cluster and whose extragalactic
origin also cannot be discarded (Contreras Pe\~na et al. 2018), NGC 6558 which is an
Oo~I of the old halo pupulation, and NGC~4590 an Oo~II type cluster with a red HB  for
its [Fe/H] and which has been considered a relatively young cluster (Chaboyer et al.
1996). The period-amplitude diagram for NGC~4590 is an interesting one (Kains et al.
2015) since the distribution of RRab stars hints the possibility of an Oo-Int
classification.

\subsection{Distribution of pulsating modes in the HB}
\label{distHB}
NGC~4147 being a cluster of 
low metallicity, having a rather blue HB and an intermediate Oosterhoff type,
the distribution of RRab and RRc stars in the HB 
is of particular interest since a clean mode
splitting has been found in all studied Oo II type clusters (nine so far), but only in
some Oo I (four out of seven so far) (Yepez et al. 2018).

Except for the RRc V16 and the RRab V12, the distribution of RR Lyrae stars on the HB
suggest a clean segregation of the RRab and RRc stars at red edge of the
first overtone instability strip at \emph{(V-I)}$_0 = 0.45$ (Arellano Ferro et al.
2016) duly reddened in Fig. \ref{CMD}.
Given the peculiarities of V12 and V16 ($\S$ \ref{indstars}), we believe they can be
omitted and hence we can conclude that in the case of NGC 4147 the fundamental
and first overtone stars are clearly segregated in the HB.

\begin{figure}
\begin{center}
\includegraphics[scale=0.45]{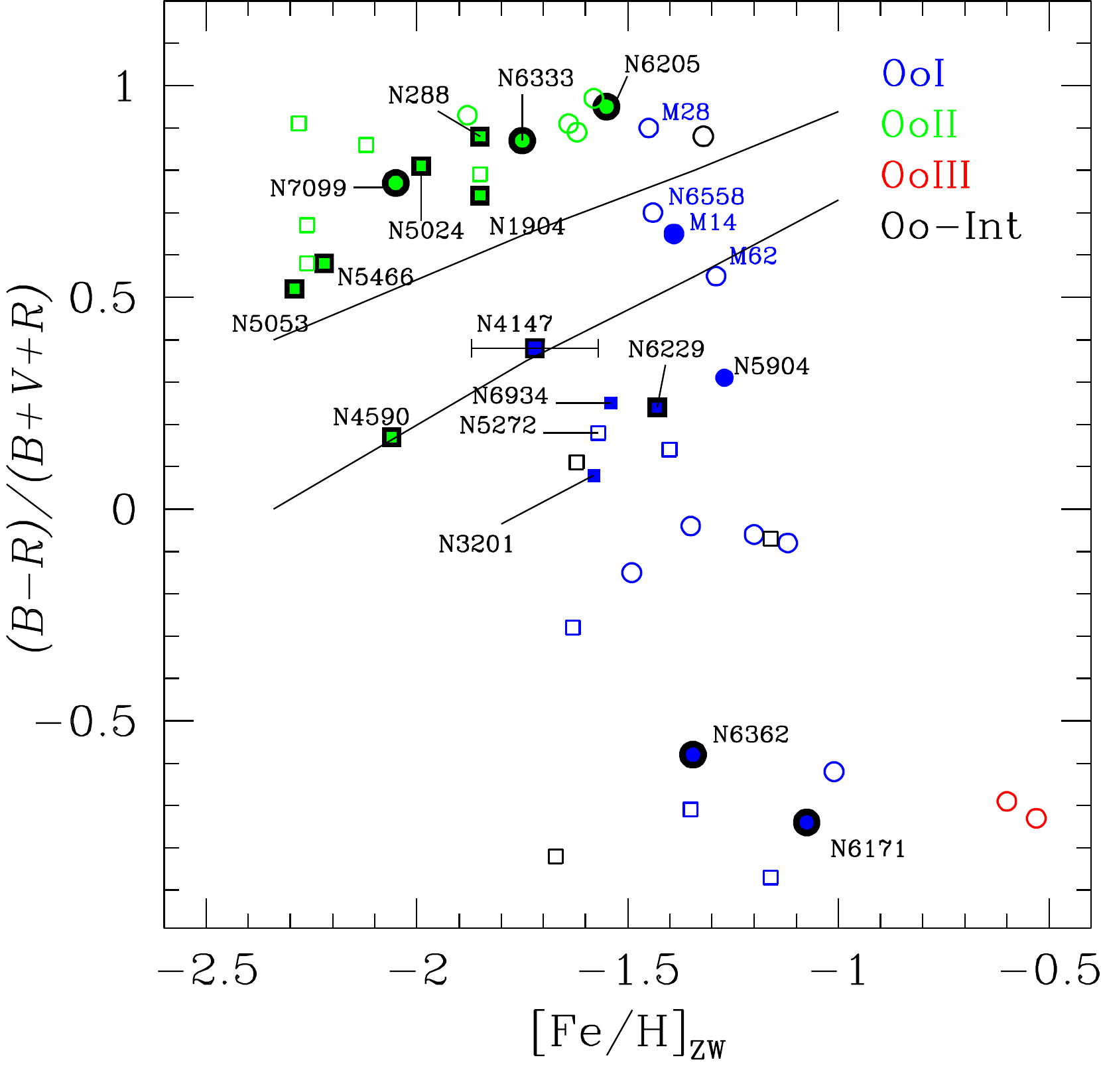}
\caption{Galactic globular clusters in the HB structure parameter $\mathcal L$ vs.
[Fe/H] plane. Circles and squares are used for inner and outer halo clusters
respectively. The black-rimmed
symbols represent globular clusters where the fundamental
and first overtone modes are well segregated around the first
overtone red edge of the instability strip, as opposed to filled non-rimmed
symbols. Empty symbols are clusters not yet studied by our
group. The case of M14 is adapted from Contreras Pe\~na et al. (2018). The upper and
lower
solid lines are the limits of the Oosterhoff gap according to Bono et al. (1994).
NGC~4147 stands out as the most metal poor Oo~I cluster but it is obviously not the
one with bluest HB in this class (see cluster labels in blue). Its position inside the
Oosterhoff gap is in consonance with the possibility of its extragalactic origin. See
$\S$ \ref{Ootype} for a discussion.}
\label{CATELAN}
\end{center}
\end{figure}

Fig. \ref{CATELAN} illustrates the occurrence of mode-splitting or mode-mixing
observed in a family of clusters of both Oosterhoff types, as observed in recent
papers of our group, on the plane HB type--[Fe/H].

As it has been discussed by Caputo et al. (1978), the mode segregation versus 
the mingled mode distribution in the inter-mode or "either-or" region is a direct
result of the mass distribution of ZAHB stars in the instability strip, which in turn
depends on the mass loss rates at the RGB (e.g. Silva-Aguirre et al. 2008). The
connection of these concepts with the so called second-parameter problem is
addressed in detail by Catelan (2009). Thus, 
it is significant that of the clusters we have revised, in all Oo~II the modes are
well segregated around the red edge of the first overtone instability strip (black
vertical line in Fig. \ref{CMD}), while this is observed only in some Oo~I. It would
be of interest to revise the RR Lyrae distributions in those clusters in the 
Oosterhoff gap and that have been associated to dwarf galaxy satellites of the Milky
Way (see Table 3 of Catelan 2009).

For the case of M14, we refer to the $B-V$ CMD of Fig 4. in the paper by
Contreras Pe\~na et al. (2018), where it is clear that the inter-mode region is
populated by both RRab and RRc stars. In passing let us point out
that NGC~4590 is the Oo~II type cluster with the reddest HB, still, its RRab and RRc
stars are well separated. 

\section{NGC 4147 in the $M_V$-[F\lowercase{e}/H] relation}
\label{MFe}

\begin{figure*}
\begin{center}
\includegraphics[scale=0.7]{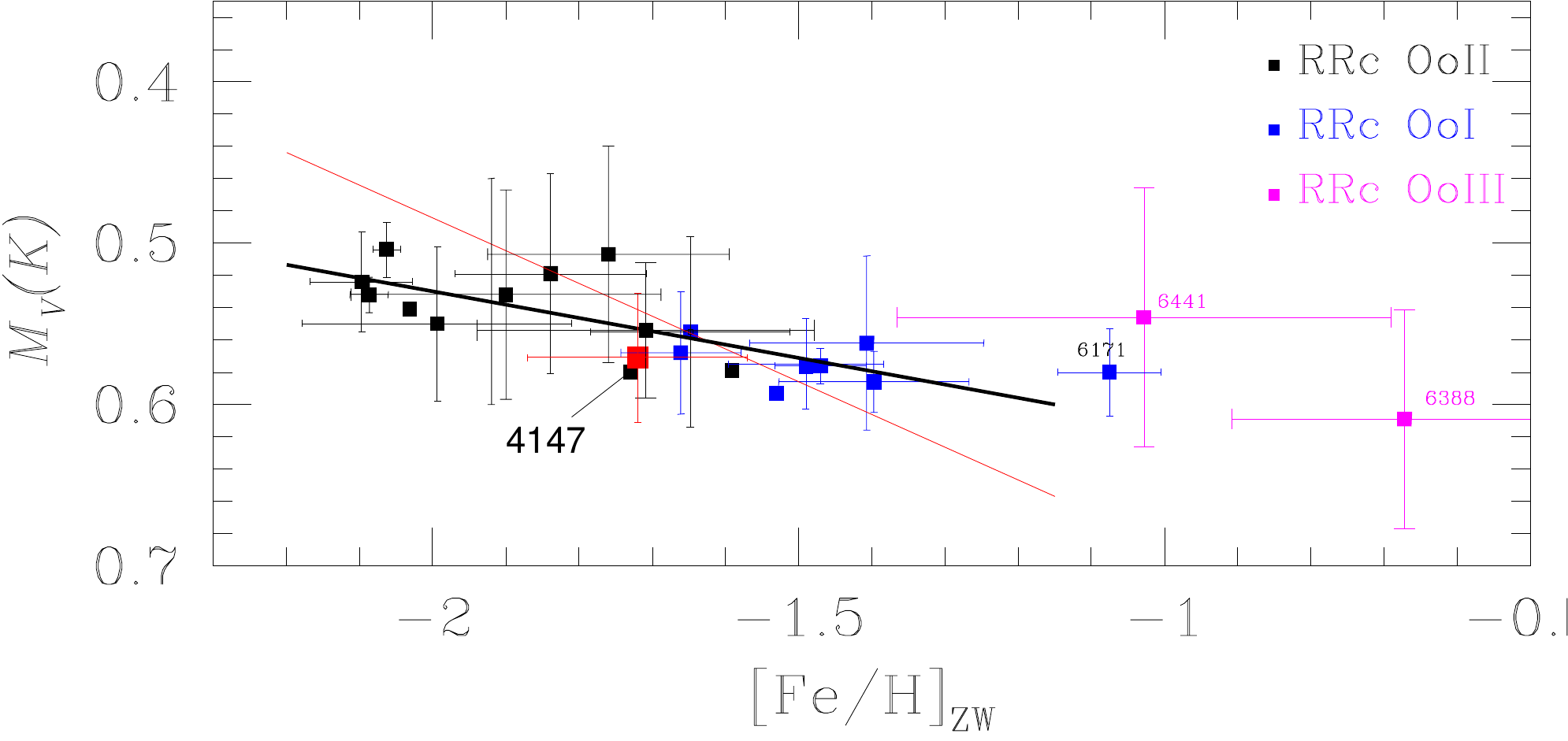}
\caption{The $M_V$-[Fe/H] relation calculated via the Fourier decomposition of RRc
star light curves. NGC 4147 does follow the general trend (thick black line) and it
stands out as the most 
metal deficient of all Oo~I globular clusters. Note the different slope obtained by a
similar analysis for RRab stars (red line) found by Arellano Ferro, Bramich
\& Giridhar 2017.}
\label{MvFe}
\end{center}
\end{figure*}

The dependence of the HB luminosity on the metallicity of a globular clusters is
now a well known fact from theoretical and empirical investigations
made over 1990's, (e.g. Lee, Demarque \& Zinn 1990, Walker 1992, Carney
et al. 1992, Sandage 1993, Chaboyer 1999). In a recent paper (Arellano Ferro, Bramich
\& Giridhar 2017) we have studied the $M_V$-[Fe/H] relation for families of Oo~I and
Oo~II groups in a homogeneous way, from the results of the Fourier decomposition of
the RR Lyrae light curves. We found that the slope of the relationship obtained from
RRab stars is significantly different from that of the RRc stars. In Fig. \ref{MvFe}
we
reproduce the
$M_V$-[Fe/H] relation for the RRc stars and place NGC 4147 using our present results
for the average $M_V$ and [Fe/H]. These values are consistent with the family of
Oo~I clusters and do follow the general trend defined by the results from the RRc
stars. Note the higher slope of the relationship when calculated from RRab star (red
line in Fig \ref{MvFe}) determined by Arellano Ferro, Bramich \& Giridhar (2017).

\section{Summary of results}
\label{sec:Summ}

New differential CCD photometry of NGC 4147 in the \emph{VI}-bands allowed us to
revise the light curves of the RR Lyrae stars in the cluster, to discuss certain
peculiarities of individual stars and calculate the mean distance and
metallicity of the cluster. The Blazhko nature of V6 and its period of about 67.46d
are confirmed. Erratic phasing in the RRc type stars V13 and V16, previously noticed
by AF04 and SCS05, is also seen in the present data. It
is suggested that it may be due to the presence of a secondary not-yet-identified
frequency similar to case of V37 in NGC 6362 (Smolec et al. 2017; Arellano Ferro
et al. 2018). 
The controversy surrounding the variable status of V18 has been resolved
now with the star clearly classified as SR variable with period about 24.8d. The
star, near the TRGB, can be used to estimate the distance to the
cluster.

From the Fourier decomposition of RRc stars we calculate the averages 
of the metallicity and distance of the parental
cluster as [Fe/H]$_{ZW}=-1.72\pm0.15$ and 19.05$\pm$0.46 kpc. It was argued that none
of the RRab stars in the cluster is suitable for the
calculation of physical quantities via the Fourier decomposition since their
light curves present strong amplitude modulations.  With the estimated metallicity,
and its HB structure parameter $\mathcal L = +0.38$,
NGC 4147 marginally falls in the Oosterhoff gap which may support its classification
as Oo-Int and its possible extragalactic origin.

The CMD displays some peculiarities. The isochrone for $\sim$[Fe/H]$_{ZW}=-1.72$,
distance and reddening estimated in this work falls to the red of the RGB by as much
as 0.05 mag, suggesting perhaps a lower metallicity. The blue tail of the HB is found
to
be too red by about 0.1 mag relative to theoretical predictions. While we do not
really have an explanation for these empirical results, we stress that the
red edge of the first-overtone instability strip falls where it is expected 
according to the results from similar analysis of other clusters. 
We found that the RRab and RRc stars 
are separated by this locus as it seems to be the case in all Oo~II but only in
some Oo~I clusters.

\vskip 1.0cm

\noindent
AAF acknowledges the support from DGAPA-UNAM grant through project
IN104917. FCRG is indebted to the Instituto de Astronom\'ia, UNAM, for warm
hospitality.
We have made an extensive use of the SIMBAD and ADS services, for which we are
thankful.

\end{document}